\documentclass{PoS}

\title{Shock acceleration of relativistic particles in galaxy-galaxy
  collisions}

\ShortTitle{Shock acceleration in galaxy-galaxy collisions}

\author{\speaker{Heinrich J. V\"olk}$^a$ and Ute Lisenfeld$^b$\\
 \llap{$^a$} Max-Planck-Institut f\"ur Kernphysik,  Heidelberg, Germany\\
\llap{$^b$} Departamento de F\'isica Te\'orica y del Cosmos, and
           Instituto `Carlos I' de F\'\i sica Te\'orica y Computacional,
           Universidad de Granada, Spain \\
           E-mail: \email{Heinrich.Voelk@mpi-hd.mpg.de}; \email{ute@ugr.es}}

  \abstract{All galaxies without a radio-loud AGN follow a tight
        correlation between their global FIR and radio synchrotron
        luminosities, which is believed to be ultimately the result of the
        formation of massive stars. Two face-on colliding pairs of galaxies,
        UGC12914/5 and UGC 813/6 deviate from this correlation and show an
        excess of radio emission which in both cases originates to a large
        extent in a gas bridge connecting the two galactic disks. The radio
        synchrotron emission expected from the bridge region is calculated,
        assuming that the kinetic energy liberated in the predominantly gas
     dynamic interaction of the respective interstellar media (ISM) has
        produced shock waves that efficiently accelerate nuclei and electrons
        to relativistic energies. A simple model for the acceleration of
        relativistic particles in these shocks is presented together with a
        calculation of the resulting radio emission, its spectral index and the
        expected high-energy gamma-ray emission. This process is not related to
        star formation. It is found that the nonthermal energy produced in the
        collision is large enough to explain the radio emission from the bridge
        between the two galaxies. The calculated spectral index at the present
        time also agrees with the observed value. The expected gamma-ray
        emission, on the other hand, is too low by a factor of several to be
        detectable even with foreseeable instruments like CTA.}

\FullConference{25th Texas Symposium on Relativistic Astrophysics \\
		 December 6-10, 2010\\
		 Heidelberg, Germany}

\begin{document}

\section{Introduction}

The universal, tight correlation between the spatially integrated far-infrared
(FIR) luminosities and the monochromatic radio continuum emissions from late
type galaxies without a bright, radio-loud AGN (de Jong et al. 1985, Helou et
al. 1985) has been known for a long time. It is believed to be ultimately a
result of the formation of massive stars and also holds for most interacting
galaxies. It is therefore quite unusual that Condon et al. (1993), hereafter
referred to as CHS, and Condon et al. (2002) found two clear exceptions to this
rule. This concerns two face-on colliding spiral galaxy systems where at the
present time, presumably some 30 Myr after the interaction, the respective
pairs of galaxy disks are again well separated from each other optically, but
are connected by a radio continuum-bright ``bridge'' of gas, suggested to be
stripped from the interpenetrating disks. The two systems show overall a
significant excess, by a factor of about two, of the radio continuum emission
relative to the FIR-radio continuum ratio expected from the FIR-radio
correlation for single galaxies. CHS interpreted the finding as the result of
general electron escape from galaxies, in this case into the bridge connecting
the pair.

In the present paper the dynamical effects of such galaxy-galaxy collisions on
the interstellar gas are investigated. It is argued that the interstellar media
of the respective galaxies will undergo a largely gas dynamic interaction,
where the low-density parts exchange momentum and energy through the formation
of large-scale shock waves in the supersonic collision. A simple model for the
acceleration of relativistic particles is presented and the synchrotron
emission from the relativistic electron component is calculated, as well as the
expected gamma-ray emission from relativistic nuclei and electrons. The model
is shown to be able to explain the radio continuum emission observed from the
bridge between the galaxies. For details, see Lisenfeld \& V\"olk (2010),
hereafter referred to as LV.

%........................

%Other recent investigations show that the FIR-radio correlation extends to even

%quite high redshifts $z<5$ \citep[e.g.][]{sargent10}. Our Milky Way galaxy is

%also argued to be an approximate electron calorimeter \citep{strong10). In

 % starburst galaxies the gas density can be so high that inelastic collisions

 % of relativistic nuclear particles may make even this component behave

 % calorimetrically \citep{aha05,thompson06}. A recent discussion of the

 % theoretical basis for the FIR-radio correlation is contained in

 % \citet{lacki10}.

%....................................

\section{Further properties of the colliding systems}

The spectral index of the radio emission between 1.49 and 4.9 GHz steepens
gradually from the stellar disks with values of 0.7--0.8 to values of 1.3--1.4
in the middle of the connecting gas bridge. This steep spectral index will be
argued to be indicative of dominant synchrotron and inverse Compton losses
suffered by the relativistic electrons. Apart from the sychrotron emitting
electrons the bridge in the system UGC 12914/5 contains also large amounts of
atomic (CHS) and molecular (Braine et al. 2003) gas, where
practically no star formation is taking place. This is interpreted as an
essentially complete hydrodynamic removal of the more diffuse atomic and
molecular gas from the galaxies. Most likely, only the dense cloud cores --
capable of forming stars -- have remained within the stellar disks together
with the stars. The other system, UGC 813/6, was described in the later paper
by Condon et al. (2002). Since it is very similar, the discussion here will be
limited to UGC 12914/5.

% ------------------------------------------------------

\section{Acceleration model}

In a face-on collision the stellar disks interpenetrate each other without
being too much altered. However, the diffuse gas and part of the gas clouds
interact hydrodynamically and exchange energy and momentum.

If one assumes that half of the gas, which is now present in the bridge, was
previously in one galaxy, and the other half in the other galaxy, then the
total energy liberated in a fully inelastic interaction is the kinetic energy
of the gas mass:

\begin{equation}
E_{\mathrm kin}  = \frac{1}{2}  M_\mathrm{gas} \left(\frac{v_\mathrm{coll}}{2}\right)^2,
\end{equation}
where $M_\mathrm{gas}$ is the total gas mass in the bridge and
$v_\mathrm{coll}$ denotes the velocity difference of the gas at collision. The
factor 1/2 converts this velocity to the velocity difference in the center of
mass system (assuming that both gas disks are equally massive). The relative
velocity between the galaxies, $v_{\rm coll} \approx 600$~km/s, in the case of
UGC 12914/5 has been derived by CHS from a analysis of the HI line and of the
galaxy masses. $M_\mathrm{gas}$ and $E_\mathrm{kin}$ are estimated as
$(4-11)\times 10^9~M_{\odot}$ and therefore $(0.49-1.4)\times 10^{58}$~erg,
respectively.

%\footnote{For a detailed derivation ofTexas_rev5.dvi  these numbers and of other
%  parameters in the following, see LV.}.

At the collision of the two ISM, a tangential discontinuity will form and two
strong shocks will propagate in opposite directions with velocities
$v_\mathrm{shock}$, communicating the interaction to larger and larger
fractions of the colliding interstellar gas masses. The space in between these
shocks is filled with post-shock gas. Fig. 1 shows the idealized picture of
this interaction, the basis of the present model, in the reference frame of the
motion normal to this tangential discontinuity, which is also the center of
mass system. In this frame, the post-shock normal velocity $v_\mathrm {post}$
vanishes and the preshock normal velocity of the gas is $v_\mathrm
{pre}=\frac{1}{2}v_\mathrm{coll}$.  The contact discontinuity is stationary and
situated in the middle between the galaxies (at $x=0$ in Fig. 1). In a strong,
but approximately unmodified, adiabatic gas shock, the normal component of the
velocity difference between the shock in the pre- and postshock gas follows the
relation: $ 4 \times (v_\mathrm{post}-v_\mathrm{shock}) =
   (v_\mathrm{pre}-v_\mathrm{shock})$.
 \noindent With $v_\mathrm{post}=0$, this yields
 $v_\mathrm{shock}=\frac{1}{6}v_\mathrm{coll}$.  The shock velocity
 $V_\mathrm{s}$, relative to the unperturbed ISM gas, is then $V_s =
 v_\mathrm{shock} + \frac{1}{2}v_\mathrm{coll} = \frac{2}{3} v_\mathrm{coll}$ =
 400 km/sec.
From Fig. 1 in CHS one can estimate that the shocks are at present close to the
 galaxy disks so that practically the entire bridge is expected to be filled
 with post-shock gas.

 The Mach number of the shocks produced in this collision is like in
 middle-aged supernova remnants in the Sedov phase. Roughly speaking, the
 particle acceleration efficiency of such shocks will therefore be
 similar to that of a supernova remnant, i.e. of the order of $10 - 30 \%$~
 (e.g. Berezhko \& V\"olk 1997). This is a basic assumption for the present
   paper.

   The source function $Q(E)$, i.e. the number of relativistic particles
   produced by the shock per energy and time interval is given by: $Q (E) = 2
   f_\mathrm{acc}(E)~v_\mathrm{shock}~A$, where $A$ is the area covered by the
   shock (roughly the area of the galaxy disks).  The factor 2 is due to the
   fact that the shocks propagate into two, opposite directions.  $
   f_\mathrm{acc}(E)$ is the downstream, uniform number of relativistic
   particles of rest mass $m$, produced per volume and energy interval at the
   shocks:

 \begin{equation}
f_\mathrm{acc}(E) = f_0  \biggl({E\over m c^2}\biggr)^{-\gamma}.
\end{equation}
Here, $\gamma$ is the spectral index of the differential relativistic particle
source spectrum, taken to be $\gamma=2.1$ (Drury et al. 1994, Berezhko \&
V\"olk 1997).  For the nuclear particles, essentially protons, the constant
$f_0$ can be determined by requiring that the total energy converted into
relativistic particles during the entire duration of the interaction, $T$, is
equal to $E_\mathrm{acc}= \Theta \times E_\mathrm{kin}$, where $\Theta =
0.1-0.3$:

\begin{displaymath}
  E_\mathrm{acc}  =   \int_{E_{\rm min}}^\infty E\, Q(E)\, T\, dE 
  =  \int_{E_{\rm min}}^\infty 2\,E f_0  \biggl({E\over m_\mathrm{p}c^2}\biggr)^{-\gamma}  v_\mathrm{shock}~
  A\,  T\,  dE 
  =   2 f_0 {(m_\mathrm{p}c^2)^2 \over \gamma-2}  \biggl({E_{\rm min}\over m_\mathrm{p}c^2}\biggr)^{-\gamma+2} 
  v_\mathrm{shock}~ A\,  T,
%  \nonumber \\
\end{displaymath} 
for $\gamma >2$. Here, $E_\mathrm{min}$ is the minimum energy of relativistic
protons of mass $m_\mathrm{p}$ accelerated, taken as $E_\mathrm{min}= 1$ GeV
(Drury et al. 1994). This gives

\begin{equation}
 f_0  = \frac{E_\mathrm{acc}}{ 2 T \, A\, v_\mathrm{shock} } 
 \frac{(\gamma-2)}  {(m_\mathrm{p}c^2)^2} 
\end{equation}

In the Galactic cosmic rays, at a given energy, the number of relativistic
electrons is about 1\% of that of the protons at GeV energies, and the source
spectra for electrons and protons are probably similar (M\"uller
2001). Assuming that this electron-to-proton ratio is also representative for
the source spectra in the galaxies considered here, implying that the electron
and proton source spectra are the same, for the source function of the
relativistic electrons $Q_{\rm e}(E) = Q(E) \times 0.01$ is adopted, which
means that their source distribution is $f_\mathrm{acc,e} =
f_{0,\mathrm{e}}~(E/(\mathrm{m_\mathrm{p}} c^2))^{-\gamma}$, with
$f_{0,\mathrm{e}} = f_0 \times 0.01$.

\begin{figure}
\includegraphics[width=7.cm, angle=90]{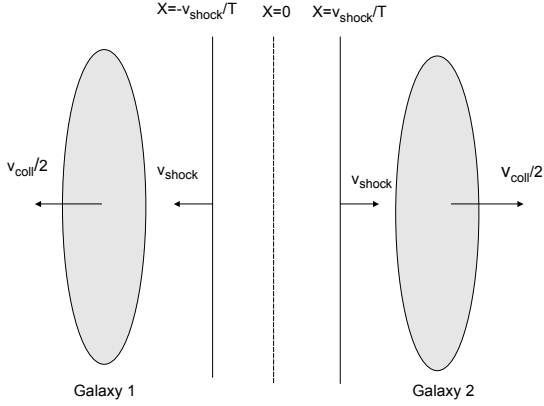}
\caption{Schematic illustration of the shock in the bridge region, in the
center of mass system.}
\end{figure}

%--------------------------------------
\section{Relativistic electron density and  synchrotron emission}

In order to calculate the synchrotron emission from the bridge the
time-dependent propagation equation for the electron particle density
$f_\mathrm{e}(t,x,E)$ is solved. Due to the spatial symmetry of the situation,
a one-dimensional approximation is appropriate, where $x$ is the coordinate in
the direction along which the galaxies separate (see Fig. 1). In addition, the
diffusion of relativistic electrons is neglected because the typical spatial
scales which are relevant on the time-scales discussed here, $3 \times 10^7$
yr, are $ < 1$~ kpc, whereas the width of the bridge is 10 kpc. Then one can
write:

\begin{equation}
{\partial f_\mathrm{e}(t,x,E)\over \partial t}
        =  q_{\rm e}(t,x,E) +
    {\partial\over \partial E} \biggr\{b
     \, E^2 \, f_\mathrm{e}(t,x,E)\biggr\},
\end{equation}
where 
\begin{equation}
  q_{\rm e}(t,x,E) = \frac{Q_{\rm e}}{A} \bigl(\delta(x-v_\mathrm{shock}t)+ 
\delta(x+v_\mathrm{shock}t)\bigr)
% \label{source}
\end{equation}
is the local source strength (in units of relativistic electrons produced per
energy interval per time and per volume). This source strength describes two
shocks that start at $t=0$ at $x=0$ and propagate into opposite directions with
velocity $v_\mathrm{shock}$.  Eq.(4.1) takes into account the electron
acceleration in the shocks together with the radiative energy losses of these
CR electrons due to Inverse Compton and synchrotron losses; up to the present
epoch, adiabatic losses can be disregarded. Then:
\begin{equation}
\left(\frac{\mathrm{d}E}{\mathrm{d}t}\right)_\mathrm{rad}  =  -bE^2 = -
\frac{4}{3} \sigma_\mathrm{T} c \left(\frac{E}{m_\mathrm{e}c^2}\right)^2
(U_\mathrm{B} + U_\mathrm{rad}), 
%& = & 2.5 \,  10 %^{-18}\left(\frac{B}{\mu \mathrm G}\right)^2
%\left(\frac{E}{\mathrm{GeV}}\right)^2 \mathrm{GeV s^{-1}},
%\label{eqeight}
\end{equation}
where $\sigma_\mathrm{T}$ is approximated by the Thompson scattering cross
section, $B$ the magnetic field strength, $U_\mathrm{B}$ its energy density and
$U_\mathrm{rad}$ is the energy density of the radiation field.
Neglecting adiabatic losses is consistent with the fact that the distribution
of the synchrotron spectral index $\alpha(1.49,4.86)$, between 1.49 and 4.86
GHz, has an integrated value $\alpha(1.49,4.86) > 1$~in the bridge region (CHS),
indicative of dominant Inverse Compton and synchrotron losses.

The magnetic field can be estimated from the minimum energy requirement to be
about 7~$\mu$G in the bridge of UGC~12914/5 (CHS), which implies $U_\mathrm{B}
= 1.2$~eV cm$^{-3}$. The total radiation field density from (in essentially
equal amounts) the blue magnitude, the FIR, and the Cosmic Microwave Background
(CMB) amounts to $U_\mathrm{rad} \approx 0.76$~eV cm$^{-3}$ (LV).

\noindent The solution of eq. (4.1)  is:
\begin{eqnarray}
  f_{\mathrm{e}}(E,|x|,t)  =  f_{\rm 0,e}
 \biggl({E\over \mathrm{m}c^2}\biggr)^{-\gamma}
% \biggl\{
  \left\{ 1-E\,b \left(t-\frac{|x|}{v_\mathrm {shock}}\right)
  \right\}^{\gamma-2}, \\ \nonumber
\hspace{5.cm} \mbox{ for } ~  t > |x|/v_\mathrm{shock}  \mbox{ and }
t-|x|/v_{\mathrm{shock}} < \tau_{\mathrm{loss}}
 \end{eqnarray}
and
\begin{equation}
f_\mathrm{e}(E,|x|,t) = 0, \mbox{ for }~t < |x|/v_\mathrm{shock} \mbox{ and }
  t-|x|/v_\mathrm{shock} > \tau_\mathrm{loss},
\end{equation}
with $\tau_\mathrm{loss} = (Eb)^{-1}$~ being the life-time of a relativistic
electron against radiative energy losses.

\noindent This expression can also be integrated over the volume of the bridge
to obtain the total number of relativistic electrons in the bridge
$F_\mathrm{e}(E,t)$. The result is:
\begin{equation}
  F_\mathrm{e}(E,t) = 2\, A\,f_{\rm 0, e} 
  \biggl({E\over\mathrm{m_\mathrm{p}}c^2}\biggr)^{-\gamma} v_\mathrm{shock}
  \tau_\mathrm{loss}\frac{1}{\gamma-1}
  \left\{1-(1-t/\tau_\mathrm{loss})^{\gamma-1} \right\}, \mbox{ for } t< \tau_\mathrm{loss} 
%\label{int_dist_f}
\end{equation}
and 
\begin{equation}
F_\mathrm{e}(E,t) = 2\, A\,f_{\rm 0,e} \biggl({E\over
   \mathrm{m_\mathrm{p}}c^2}\biggr)^{-\gamma} v_\mathrm{shock}
   \tau_\mathrm{loss}\frac{1}{\gamma-1}, \mbox{ for } t>\tau_\mathrm{loss}
\end{equation}
\noindent The synchrotron spectrum is obtained by convolving $f_\mathrm{e}(E,x,t)$,
respectively $F_\mathrm{e}(E,t)$, with the synchrotron emission spectrum of a
single electron.

This solution allows a theoretical prediction for the total radio flux density
specifically at $1.49$~GHz and of the radio spectral index between $1.49$ and
$4.86$~GHz. Both values are in satisfactory agreement with the observations
(LV). Clearly also the radio spectral index increases towards the center of the
bridge. The model calculations therefore show that acceleration by large-scale
shocks caused by the galaxy-galaxy interaction is indeed able to explain the
radio emission from the bridge both in intensity and spectral morphology.

The two interacting galaxy pairs studied here, are possibly examples of what
might have happened much more frequently at early stages of structure
formation, when primordial galaxies had already developed magnetic fields as a
consequence of early star formation, but when they were still likely to
interact strongly with neighboring structures of a similar character.

\section{High-energy gamma-ray emission}

Although not the primary topic of this paper, it is clear that the interaction
of galaxies considered here will also lead to the acceleration of gamma-ray
producing very high-energy particles, both nuclei and electrons, in the form of
the distribution $f_\mathrm{acc}$, cf. eq. (3.2). The visibility of the
acceleration process also in high-energy gamma rays would be an independent
argument for the model presented. In the following a rough estimate will be
given.

The shock system which characterizes the face-on interaction of the two spiral
galaxies can to first approximation be considered as plane parallel, with
constant speed shocks; the energetic particles remain confined in the growing
interaction region until the interaction is completed. Assuming the diffusion
coefficient to be as low as the Bohm diffusion coefficient (e.g Kang 2007) one
can approximately calculate the maximum proton energy achieved at the present
epoch as $E_\mathrm{max} \approx 3.7 \times 10^{16}$~eV. This very high energy
is the result of the long life-time ($T=2.8 \times 10^7$~yrs) of the shocks,
inspite of their comparatively moderate speed of $\approx 400$~km
sec$^{-1}$. The situation is rather different for the energetic electrons
because of their radiative losses. Even disregarding any magnetic field
amplification at the shocks due to the accelerating particles, the maximum
electron energy is only $\sim 10$~TeV. For the IC gamma-ray emission in the
sub-TeV region mainly the CMB counts. In contrast to the great majority of the
radio electrons, however, all gamma-ray emitting electrons are ``old'' as a
result of postshock radiative losses. As a consequence their energy spectrum is
softened with a correspondingly reduced IC gamma-ray emission.

When the shocks have gone through the interacting ISM of the two galaxies,
which is the case at about the present epoch, the kinetic energy
$E_\mathrm{kin} = (0.49 \mbox{ to } 1.4)\times 10^{58}$~erg has been
transformed into thermal and nonthermal particle energy. This corresponds to
$E_\mathrm{acc} \sim 10^{57}$~erg, predominantly in relativistic nuclei, which
is roughly $10^7$ times more energy than available from a single supernova
remnant.

The bridge volume $V$ is reasonably estimated as $V \approx \pi R^3$, where the
disk radius $R \approx 10$~kpc (CHS). This implies a mean gas
density of $0.05 \mbox{ to } 0.14$~cm$^{-3}$, which is an order of magnitude
smaller than the typical density in the plane of spiral galaxies and more than
three orders of magnitude smaller than the density in the starburst
nucleus of e.g. the galaxy NGC 253.

An analytical estimate for the integral hadronic gamma-ray emission, from
$\pi^0$-production by collisions of energetic protons with gas nuclei and
subsequent decay into two gamma rays, is given in Eq.~(9) of Drury et al. (1994) for gamma
energies $E$ large compared to 100 MeV:

\begin{displaymath}
F( > E, t=T)  \approx  9 \times 10^{-11} \Theta \left(\frac{E}{1 \mathrm{TeV}}\right)^{-1.1}
\left(\frac{E_\mathrm{kin}}{10^{51}\mathrm{erg}}\right) \left(\frac{\mathrm{d}}{1
\mathrm{kpc}}\right)^{-2} \times \left(\frac {n}{1~\mathrm{cm}^{-3}}\right )
\mathrm{photons}~ \mathrm{cm}^{-2}~ \mathrm{s}^{-1}
% \label{gamma_general}
\end{displaymath}
\noindent Inserting the values $E_\mathrm{kin} = 10^{58}$~erg, $d=61~\mathrm{Mpc}$,
$\Theta =0.1$, and $n=0.05 \mbox{ to } 0.14$~cm$^{-3}$ results in
\begin{equation}
(-1) E^2 {\partial F( >E)\over \partial E}  \approx  (2 \mbox{ to } 6) \times 10^{-15} 
%  \nonumber \\
 \mathrm{erg}~ \mathrm{cm}^{-2}~ \mathrm{s}^{-1},
%\label{gamma_hadronic}
\end{equation}
approximately independent of energy up to about $10^{15}$~eV.

For the IC emission of the electrons at high gamma-ray energies the
distribution $F_\mathrm{e}(E,t)$ for $ t>\tau_\mathrm{loss}$ in eq. 4.7 is
relevant. For $\gamma \approx 2$ the spectral energy flux density is roughly
energy independent with a value $ \sim 7 \times
10^{-15} \mathrm{erg}~ \mathrm{cm}^{-2}~ \mathrm{s}^{-1}$.  

Thus the IC gamma-ray energy flux is of the same order as the hadronic
gamma-ray energy flux in the region of energy overlap, even for the radiatively
cooled electrons. This is a consequence of the low mean gas density in the
bridge. On the other hand, the lowest TeV photon flux from an astrophysical
source detected until now was { $F( >220~\mathrm{GeV}) = 5.5 \times 10^{-13} \,
  \mathrm{photons}~ \mathrm{cm}^{-2}~ \mathrm{s}^{-1} $. } For a flat spectral
energy density this corresponds to $8.8 \times 10^{-13}~\mathrm{erg}~
\mathrm{cm}^{-2}~\mathrm{s}^{-1}$. The measurement was made with the
H.E.S.S. telescope system for the nearby starburst galaxy NGC253 (Acero
2009). Taking this result as a yard stick, the expected hadronic flux from UGC
12914/5 at gamma-ray energies above {1 TeV} is still two orders of magnitude
below this minimum flux. The expected minimum detectable energy flux for the
future {\it Cherenkov Telescope Array} (CTA) is as low as $\approx 5 \times
10^{-14}$~erg~cm$^{-2}$~s$^{-1}$ in the TeV region, and at least one order of
magnitude higher at 50 GeV (CTA 2010). Therefore the gamma-ray flux is also
below the detection capabilities of CTA. To this extent the more optimistic
expectation by LV is corrected here. This flux is also expected to be too low
for the detection capabilities with the LAT instrument on {\it Fermi} at lower
gamma-ray energies. The same source at the distance of NGC 253 would be
detectable even for present Northern Hemisphere ground-based TeV gamma-ray
instruments like VERITAS and MAGIC, and at GeV energies for {\it
  Fermi}. Ultimately the reason for the low gamma-ray fluxes from these distant
interacting systems is their low gas density and their comparatively high age,
despite the large overall nonthermal energy they contain.

\end{document}